\documentstyle[12pt,epsfig]{article}

\def\thebibliography#1{\section*{REFERENCES}\small \list{\arabic{enumi}.}
  {\settowidth\labelwidth{#1.}\leftmargin=1.67em
   \labelsep\leftmargin \advance\labelsep-\labelwidth
  \itemsep 0 pt \parsep 0 pt
   \usecounter{enumi}}\def\makelabel##1{\rlap{##1}\hss}%
   \def\newblock{\hskip 0.11em plus 0.33em minus -0.07em}
   \sloppy \clubpenalty=4000 \widowpenalty=4000 \sfcode`\.=1000\relax}

\begin{document}
{\  }
\noindent 
 To appear in {\it The gap Symmetry and Fluctuations in High Temperature 
 Superconductors}  Edited by J. Bok, G, deutscher, D. Pavuna and S.A. Wolf. 
 (Plenum Press, 1998).

  \vskip 0.5 cm
 \noindent
 Proceedings of NATO ASI summer school held September 1-13, 1997 in Carg\`ese,
   France. (cond-mat/9901333)
  \vskip 2.5 cm

\noindent 
\bf 
\centerline{FROM MAGNONS TO THE RESONANCE PEAK:}
\centerline{SPIN DYNAMICS IN HIGH-$T_C$ SUPERCONDUCTING CUPRATES}
\centerline{BY INELASTIC NEUTRON SCATTERING}
\vskip 1.5 cm

\rm
\parindent 1in
Philippe Bourges

\vskip .5 cm
\parbox[1 in]{5 in}{ Laboratoire L\'eon Brillouin,\\
 CEA-CNRS, CE Saclay,\\ 91191 Gif sur Yvette, France }

\parindent 0.9 cm

\vskip 1.5 cm

\noindent 
{\bf 1. INTRODUCTION}
\vskip .5 cm

Over the last decade, inelastic neutron scattering (INS) experiments 
have provided a considerable insight in the understanding of the 
anomalous properties of high $T_c$-superconductors. Neutron 
measurements have shown the persistence of antiferromagnetic 
(AF) dynamical correlations over the whole metallic state of 
cuprates\cite{rossat2} which demonstrates the strong electronic 
correlations existing 
in metallic cuprates. Together with the nuclear magnetic resonance 
(NMR)\cite{rmn,rmnhere} and later with bulk magnetic 
susceptibility\cite{batlogg}, INS has evidenced the ``spin pseudogap'' 
phenomenon in underdoped cuprates, a topic of intense current interest. 
Further, INS has shown that these spin excitations 
are very intimately linked to superconductivity as a sharp magnetic 
 peak occurs when entering the superconducting  state. This
peak,  referred to ``resonance'' since its first evidence 
by J. Rossat-Mignod {\it et al}\cite{rossat1}, has spawned a considerable 
theoritical activity.

The structural peculiarity of high-$T_c$ cuprates is that they are all 
built from stacking of ${\rm CuO_2}$ planes separated by different kinds
of layers, the ``charge reservoirs", which are essential as they control 
the charge transfer mechanism. The ${\rm CuO_2}$ plane is of central 
importance as it carries most of the anomalous 
physical properties in the normal state and, likely, 
contains the keypoint for the mechanism of high-$T_c$ superconductivity 
as it has been proposed in many different approaches (See e.g. 
\cite{model,pines,so5,ops,bok,varma}). The unusual properties of high-$T_c$ 
cuprates are extensively reviewed in this book and will not be discussed 
here in details. 

\begin{figure}
\parbox{7.5 cm} {
\hspace*{-0.5 cm}
 \epsfig{file=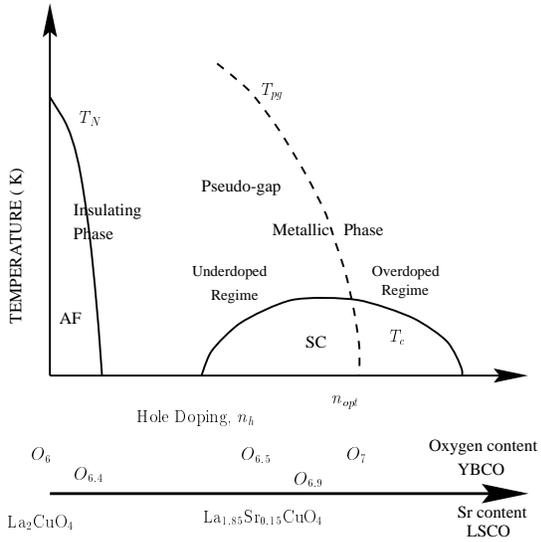,height=6.7 cm,width=7 cm} 
\caption[diag]{\label{diag} Schematic phase diagram of 
high-$T_c$ cuprates. The dashed line corresponds to a crossover 
temperature, $T_{pg}$, below which most physical properties exhibit 
or infer a pseudogap behavior. Real systems are indicated versus hole
doping in the bottom part. Note that $T_N$ in 
YBCO disappears for  $x \sim 0.4$ which 
actually corresponds to only about 2-4\% of holes in each $CuO_2$ 
plane\cite{rossat2}.}  } 
\hspace{ .2 cm}
\parbox{7.5 cm} {
\hspace*{+0.5 cm}
\epsfig{file=disp_af.epsi,height=6.7 cm,width=8.5 cm}
\caption[dispAF]{ \label{dispAF} Spin-wave dispersion  
along the (110) direction in undoped YBCO. 
Inset displays the scattering plane in the reciprocal space 
commonly used in INS measurements with $Q=(h,h,q_l)$ or
$Q=(3 h,h,q_l)$.  Squares represents nuclear Bragg peaks and circles 
 AF Bragg peaks occuring in the  N\'eel state. The two arrows 
indicate a typical Q-scan trajectory performed across the 
magnetic line. } }
 \end{figure}

A generic phase diagram, sketched in Fig. \ref{diag},
 has been established on phenomenological grounds.
Starting from an insulating and antiferromagnetically ordered state around
zero doping with one electron per Cu d$_{x^2-y^2}$-orbital, the 
cuprates become metallic by introducing holes (or electrons) from the charge 
reservoirs to the $CuO_2$ plane. A superconducting (SC) phase occurs at
further doping defining an optimal doping, $n_{opt}$ when 
the superconducting transition, $T_c$, is passing through a maximum. 
Doping rates, below and above $n_{opt}$ define usually called 
underdoped and overdoped regimes, respectively. In the underdoped state, 
many physical properties exhibit anomalous behavior below 
a crossover temperature, $T_{pg}$: that is the case for the macroscopic 
spin susceptibility, NMR Knight shift, specific heat, transport
properties\cite{batlogg}... In that doping range, it has been 
shown by NMR\cite{rmn,rmnhere} and INS\cite{rossat2,rossat1} experiments 
that the spin fluctuations are charaterized  at low temperature by 
the opening of a spin pseudo-gap. 
Recently, photoemission experiments\cite{ding,campu} have evidenced that the 
single-particle excitation spectrum exhibits also a pseudo-gap in the 
normal state below $T_{pg}$. Similar observations have been done in
optical conductivity\cite{uchida} and Raman scattering\cite{raman}
measurements. Concomitantly, these pseudo-gap observations disappear
above the optimal doping. Few attempts have been 
made to describe the phase diagram which have raised questions
such as: do the antiferromagnetic fluctuations alone explain 
the observed phases\cite{model,pines,so5}, or,  do we need to consider 
a new critical point at optimal doping which would scale the physical 
properties\cite{florapfeuty} as proposed, for instance, in the 
``circulating current'' phase\cite{varma} ?
Conversely, a comprehensive microscopic description of all these 
gap observations is still missing at present. 

Here, it is important to relate this generic phase diagram of 
cuprates to the phase diagram of the two systems on which 
INS experiments have been performed so far. In particular, 
in ${\rm YBa_{2}Cu_{3}O_{6+x}}$ (YBCO) system, the relation
between hole doping, $n_h$, and oxygen content is not obvious 
due to the charge transfer mechanism from the Cu-O chains to the
${\rm CuO_2}$ planes\cite{uimin}. Oxygen concentrations from $x \simeq 0$
to $x \simeq 1$ cover a major part of the phase diagram and
are reported on Fig. \ref{diag} for typical doping regimes which 
will be discussed here: the optimal doping being realized for $x=0.94$. 
In the ${\rm La_{2-x}Sr_xCuO_{4}}$ (LSCO) system, maximum $T_c$ is
reached for $x=0.15$ and is generally assumed to match the optimal doping.
A close inspection of the neutron results rather suggest that it could
 correspond to an underdoped regime. The superconductivity 
could be simply reduced at further doping  because of the 
proximity of structural instabilities.

Furthermore, considering the unusual properties of 
these materials, one needs to know which kind of magnetism 
is observed in INS experiments: do we observe spin dynamics associated 
with the localized copper spins or rather related to the itinerant 
quasi-particles ? These different hypotheses have been widely addressed 
in the literature. As a limiting case,  the slightly-overdoped 
YBCO$_7$ ($x \simeq 1$) could likely be described as a simple itinerant 
magnetism since, as expected for usual metals,  antiferromagnetic 
spin fluctuations are practically not sizeable in the normal 
state\cite{hoechst,tony1}. Unfortunately, this simple Fermi liquid approach 
fails at lower doping  since  dynamical AF correlations are unambiguously 
observed up to the optimal doping. One then needs to take into account 
the underlying antiferromagnetic background. Furthermore, the anomalous 
spectral lineshape detected in photoemission experiments\cite{ding} 
exhibits a behavior inconsistent with conventional band theory indicating 
that the single-particle spectra are necessarily renormalized due 
to electron-electron interactions. These ``dressed'' quasiparticles 
could even be strongly coupled to collective excitations centered at 
the AF momentum\cite{shen,norman}, namely the spin fluctuations.
The determination of the energy, momentum, doping and temperature 
dependences of the spin excitation spectrum is then of primarily 
importance to describe the anomalous properties of these materials.

Going  from well-defined magnons (in the N\'eel undoped state) related 
to the localized copper spins to a Fermi liquid picture (at the 
highest doping available), INS experiments cover a whole range of 
situations where the spin and charge responses are intimately linked. 
More generally, INS observations provide direct information about the 
electronic interactions within $\rm CuO_2$ planes, and even, between 
the two adjacent metallic $\rm CuO_2$ layers in YBCO. Here, I shall 
review the doping evolution of the spin dynamics. After a short recall
of the neutron technique (Sec. 2) and of the experimental difficulties 
(Sec. 3), magnons in the AF state are reported in Sec. 4. 
The momentum and energy dependences in the metallic state are
presented in Sec. 5 and 6, respectively. The strong modifications 
of the spin dynamics induced by superconductivity are discussed in Sec. 7.
The normal state spin susceptibility, characterized by a ``spin 
pseudogap'', is emphasized in Sec. 8. Most of the results described here
concerns the YBCO system although comparisons with the LSCO system 
are occasionally  made. 

\vskip 1 cm
\noindent 
{\bf 2. INELASTIC NEUTRON SCATTERING}
\vskip .5 cm

The interaction of neutrons with the condensed matter is 
double\cite{lovesey}: nuclear interaction with the atomic nucleus 
and magnetic dipolar interactions between the neutron 
spin and magnetic moments in solids, spin of unpaired electrons for
instance. Neutron scattering is then 
a very unique tool as it measures in the same time 
structural information (as can do X-ray scattering) but also  magnetic 
properties. Furthermore, thermal neutrons which possess a wavelength
of the order of the atomic distances, 0.5-10 \AA,  have in the same time 
an energy, 0.1 - 200 meV, which covers the large range of excitations 
in solids. Inelastic neutron scattering  then  gives invaluable 
information on both spatial and time-dependent of nuclear 
(like phonons) and magnetic correlations whose 
momentum and energy dependences are only accessible using INS. An extensive 
review of the possibilities of neutron scattering can be found 
in a recent course\cite{hercule}. Here, it is worth emphasizing that 
single crystals are required to determine a complete momentum dependence 
of  dynamical properties. Also, due to the weak interaction of neutrons 
with condensed matter, neutron scattering probes samples in bulk, but 
conversely requires large samples. 

In high-$T_c$ cuprates,  INS using the triple-axis spectrometer 
technique\cite{hercule} has been widely used to study phonons\cite{reichardt}.
Here, we focus on magnetic scattering whose cross section is directly 
proportional to the scattering function which is identified to the 
Fourier transform in time and space of the spin-spin correlation 
function\cite{lovesey} as,

\begin{equation}
{
S^{\alpha\beta}(Q,\omega)={1 \over{2 \pi\hbar}} \int^{+\infty}_{-\infty}
    dt \exp(-i\omega t) <S^{\alpha}_{Q} S^{\beta}_{-Q}(t)>.
\label{sqw}
}
\end{equation}

This scattering function is in turn related to the imaginary part of the
dynamical generalized spin susceptibility, $Im\chi(Q,\omega)$, by the 
fluctuation-dissipation theorem. Here, we consider a single component 
of the dynamical generalized susceptibility tensor associated with 
Carthesian spin coordinates $S^{\alpha}$ and $S^{\beta}$ with
 $\alpha,\beta=x,y,z$,
\begin{equation}
{
\chi^{\alpha\beta} (q,\omega)= -(g \mu_B)^2 {i\over{\hbar}} 
\int_0^{\infty} dt \exp^{-i\omega t} <[S^{\alpha}_Q(t),S^{\beta}_{-Q}]>.
\label{imchi}
}
\end{equation}

In isotropic magnetic systems, $\chi$ is simply 
identified to ${\rm Tr}(\chi^{\alpha\beta})/3$. $Im\chi(Q,\omega)$ is 
a very useful quantity,  especially when no  theory can be used to describe 
the data. It basically contains all the physical interest 
and can be calculated in different microscopic models.  
In undoped N\'eel state, the spin susceptibility describes the excited 
states above the AF ground state, referred to as magnons\cite{lovesey}, 
and neutron scattering cross-sections are well described using 
spin-wave theory in Heisenberg model. When adding a small amount of
holes in $\rm CuO_2$ planes ($n_h\le 5$\%), the system remains in the 
insulating state, but very peculiar spin dynamics is observed:
the low energy excitations are strongly enhanced at low 
temperature\cite{keimer,yvan,sendai} likely due to electron-hole interactions. 
Increasing further the doping, spin fluctuations are still detected in the 
metallic state. Further, by calibration with phonon scattering, INS 
experiments provide absolute units for the dynamical susceptibility which 
are important for theoritical models. This absolute unit calibration is 
also necessary to compare the INS results in different systems as well 
as  results obtained using different techniques (with NMR measurements 
for instance). 

\begin{figure}
 \epsfig{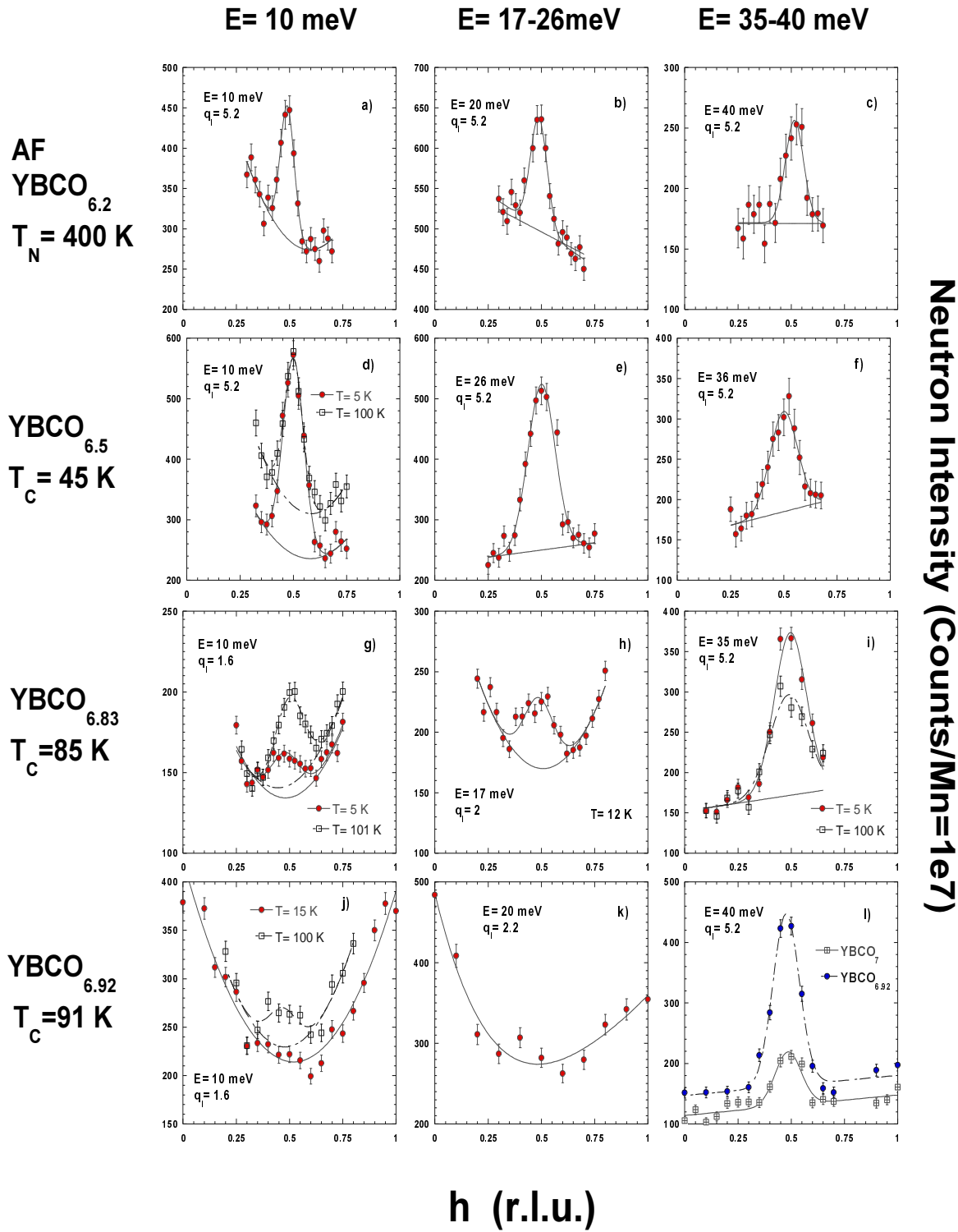}
\caption[allqscan]{Q-scans performed across the magnetic line,
$Q=(h,h,q_l)$, where $h$ is scanning over two Brillouin zones.
All neutron intensities are normalized to the same units, 
and are then directly comparable each others. The $q_l$ value along the 
(001) direction was chosen to get the maximum of the magnetic structure 
factor\cite{rossat2}. In the AF state ($x=0.2$), only a single 
peak  is  observed for the counterpropagating spin-waves; this is caused by
resolution effect of the spectrometer. Note the strong reduction of the 
E= 40 meV peak intensity from YBCO$_{6.92}$ (nearly optimally doped regime)  
to  YBCO$_{7}$ (overdoped regime) (Fig. \ref{allqscan}.l).  }
\label{allqscan} \end{figure}

In undoped cuprates, the ordered AF phase is characterized by an
in-plane propagation wave vector, $q_{AF}= ({1\over2},{1\over2},0)$ and 
the magnetic neutron cross section is maximum at wave vectors like 
${\bf Q= \tau + q_{AF}}$, where $\tau$ denotes Bragg peaks of the 
nuclear structure (squares in the inset of Fig. \ref{dispAF}). 
Roughly speaking, when going into the metallic state, the magnetic 
scattering remains always peaked around the in-plane component of the
AF wave vector. This occurs for any values of the momentum transfer 
$q_l$ perpendicular to the plane. The magnetic scattering is then concentrated
around lines in the reciprocal lattice, like $({h\over2},{k\over2},q_l)$ 
with $h$ and $k$ integers, denoted magnetic lines. Q-scans, performed 
within a Brillouin zone across these lines and at different energy 
transfers (sketched by the arrow in Fig. \ref{dispAF}), exhibit a maximun 
around $q_{AF}\equiv(\pi,\pi)$ (Fig. \ref{allqscan}) which is likely 
identified to the magnetic scattering.

\vskip 1 cm
\noindent
{\bf 3. EXPERIMENTAL EVIDENCE OF MAGNETIC EXCITATIONS}
\vskip .5 cm

Unfortunately, INS experiments are not only measuring  the magnetic 
scattering: other contributions occur either due to intrinsic nuclear 
scattering of the lattice, like phonons, or even due to spurious effects 
and impurities. Furthermore, owing to the relatively weak cross section of the 
magnetic scattering, its extraction from the total scattering 
is a major experimental problem encountered in INS experiments. 
In principle, polarized neutron beam experiments should easily 
separate these contributions,
 the magnetic scattering appearing in the 
spin-flip channel at the difference of nuclear scattering. However, 
due to the lack of statistics, polarized neutron results have yielded 
partly erroneous conclusions in high-$T_c$ cuprates \cite{dai}. 
Therefore, unpolarized neutron experiments have been largely employed. 
Due to the experimental difficulty, new results always need to be 
crosschecked and confirmed by other measurements because a single isolated 
experiment can be unfortunately misled by spurious effects. 
It has led to extensive, 
sometimes contradictory, discussions over the last 
decade\cite{rossat2,hoechst,tony1,dai,mook,sympo,epl,lpr}. 
{\it Only} the use of complementary methods as well as  the accumulation 
of neutron data allow to overcome the experimental difficulties. 
Here, I shall briefly recall the 
main guidelines which have been used to estimate the magnetic signal. 

Among these methods, let us emphasize the use of the momentum and
temperature dependences  because they are different 
for magnetism and phonons\cite{lovesey}. Especially, as a result of 
the magnetic form factor, magnetic scattering is known to decrease on 
increasing the amplitude of the wave vector over few Brillouin zones
 in contrast to the phonon scattering. Within a single 
Brillouin zone, one could also discriminate both signals by their 
q-dependences.  Further, INS experiments have 
been also performed in different scattering planes to avoid some
specific phonons\cite{tony1}.

Another powerful method developed for the YBCO system has been 
to conduct experiments in the different regimes of the phase diagram 
{\it on the same sample}\cite{rossat2,sympo,lpr}. Indeed, the 
YBa$_2$Cu$_3$O$_{6+x}$ system offers the great opportunity to cover the 
whole high-$T_c$ cuprates phase diagram just by changing the oxygen 
content by thermogravimetry from the N\'eel state, $x\simeq 0$, to the 
overdoped metallic, $x\simeq 1$. Measurements scanning the wave vector 
along the (110) direction are shown in Fig. \ref{allqscan} at few different 
fixed energy transfers for few different states of YBCO. Further, these 
experiments have been performed on the same triple axis spectrometer 
(2T-Saclay) using the same experimental setup, i.e. the same spectrometer 
resolution. This gives the great advantage that impurity 
contributions and, in a less extent, the phonons are basically 
the same in all experiments: this facilitates the extraction of the 
magnetic scattering. 

\begin{figure}
\parbox{7.5 cm} {
\epsfig{file=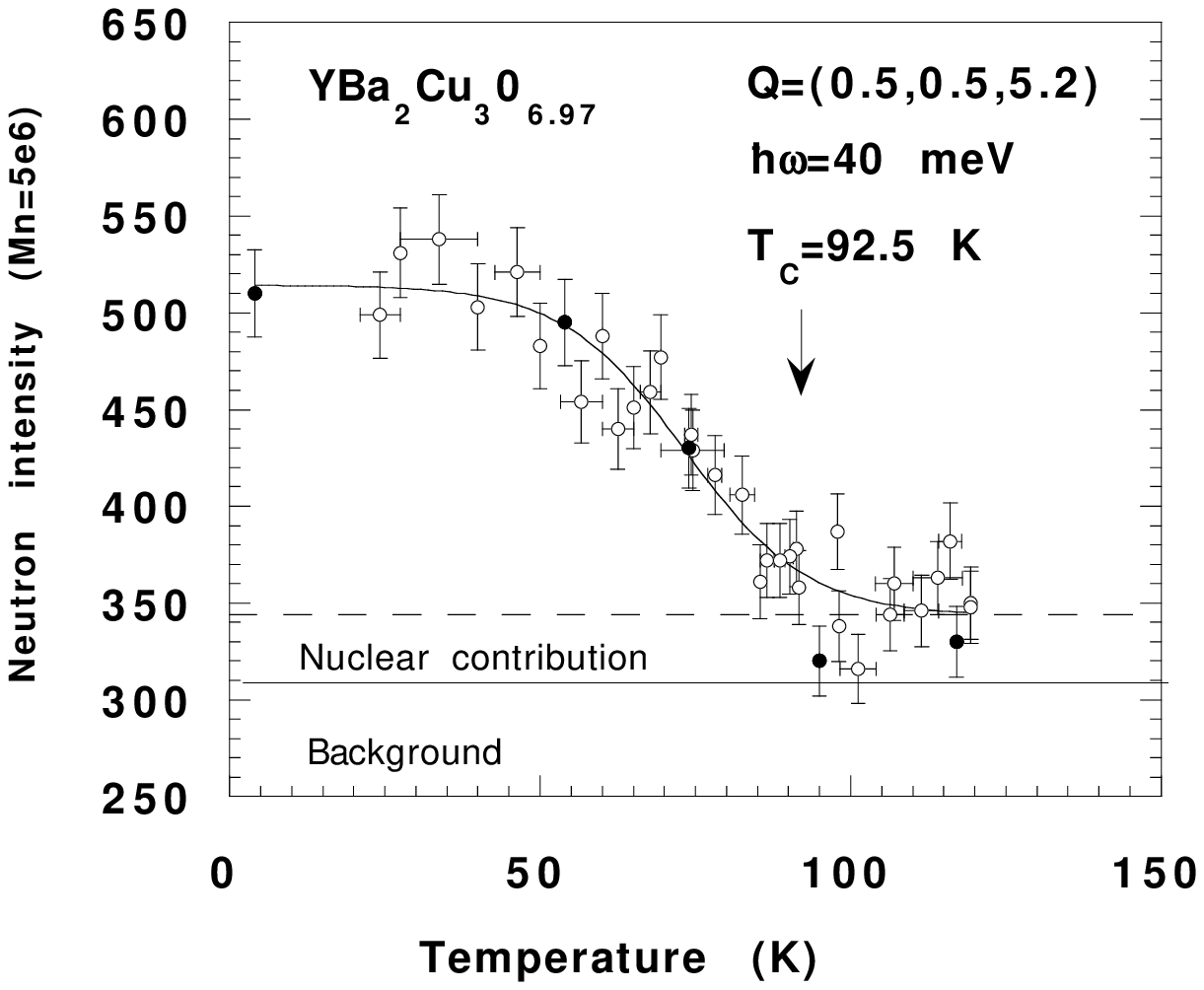,height=8 cm,width= 7 cm}
\caption[resotemp]{\label{resotemp}  Temperature dependence of the 
resonance peak at E= 40 meV (from \cite{hoechst}). The reported 
``nuclear contribution'' is due to a phonon peak whose maximum 
intensity is located at 42.5 meV\cite{tony1}.  
  }
 }
\hspace{ .2 cm}
\parbox{7.5 cm} {
\epsfig{file=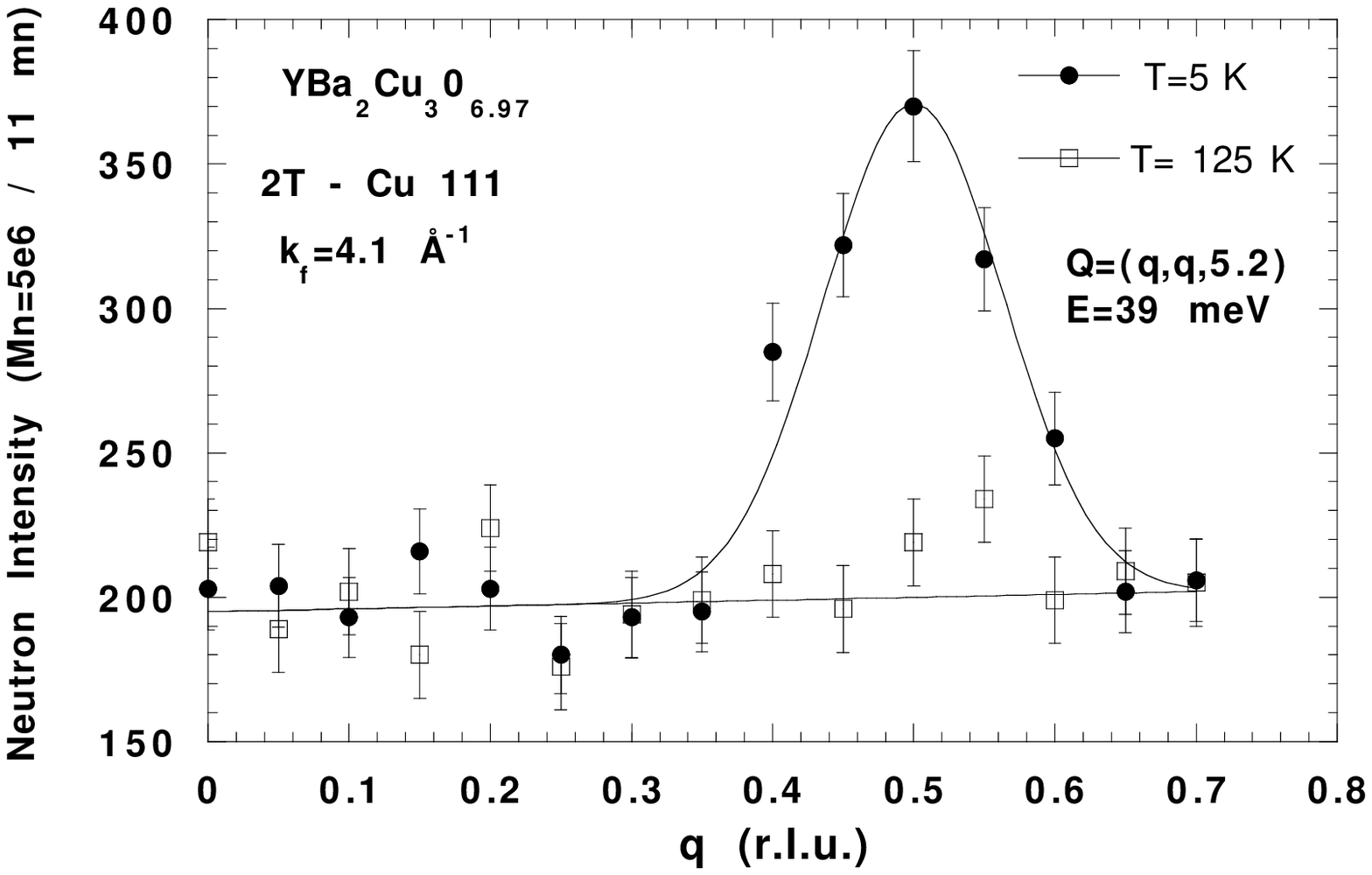,height=7 cm,width=7 cm}
\caption[qscan39]{\label{qscan39} Q-scans across the magnetic line at the 
resonance energy in the overdoped regime. These data are giving the 
most accurate upper limit of the normal state AF intensity
 which is about 8 times smaller than the resonance peak observed in 
the SC state. 
} }
 \end{figure}

These scans 
exhibit at any doping well-defined maxima at $q=0.5$, corresponding 
to the AF wave vector $\equiv(\pi,\pi)$, whose magnitude evolves with 
doping. At $E=10$ meV, a correlated scattering signal is seen at lower
doping whereas it is absent for optimally doped samples. In contrast, 
the correlated signal at $E=40$ meV first increases with the
oxygen content and then decreases for $x\simeq 1$. Measurements at nearly 
the same energy in the normal state exhibit, at most,  a very weak magnetic 
signal (see Fig \ref{qscan39}). These striking doping, energy and 
temperature dependences have given strong guidelines to analyse the data. 
In particular, two clear limits are well-defined: the undoped case where 
the  theoritical spin-wave cross section is known and the case of the 
normal state in the overdoped regime where the very weak magnetic scattering 
can be neglected. A generic shape for the non-magnetic 
contributions is then deduced for any doping. The use of such empirical 
methods, as well as phonon calculations\cite{tony2},  give a 
self-consistent picture which has made possible to improve the data analysis 
all over the years\cite{rossat2,rossat1,sympo,lpr}. When using the 
triple-axis technique, a good confidence is now reached about the 
determination of the magnetic signal over a wide range of 
energy \cite{bourges,tony4}. However, it should be mentioned 
that this method excludes contributions which are weakly momentum-dependent.
Unfortunately, no current neutron experiment is able to evidence such 
hypothetic magnetic contributions.

\vskip 1 cm
\noindent
{\bf 4. ANTIFERROMAGNETIC STATE  AND EXCHANGE PARAMETERS}
\vskip .5 cm

Undoped parent compounds of high-$T_c$ cuprates are Mott-Hubbard insulators
which are  usually described by a spin-${1\over2}$ antiferromagnetic 
Heisenberg model on a square lattice\cite{manou}. They exhibit an 
antiferromagnetic ordering  below a N\'eel temperature ranging between 
250 K and 420 K\cite{rossat2}. The most important parameter is  the 
Cu-O-Cu nearest neighbor superexchange interaction, $J$, within the 
$CuO_2$ plane.  

In ordered magnetic systems, INS experiments probe the spin-wave 
dispersion relations which relate the magnon energy, $\omega_q$, to 
the scatterred wave vector (see Fig. \ref{dispAF}). At sufficiently 
low energy (but notably above the small magnon gaps related to 
interlayer coupling and exchange anisotropies\cite{rossat2,shamoto}), the 
acoustic magnon dispersion relation starts linearly in AF systems as 
$\omega_q = c q$ (Fig. \ref{dispAF}).  $J$  is then deduced 
from the measured spin-wave velocity $c$, as $c= 2 S {\sqrt 2} Z_c J a$ 
(where $a$ is the square lattice constant, S=${1\over2}$ and $Z_c\simeq 
1.18$ represents quantum corrections of the AF ground state). 
Unfortunately, because of the large value of $J$ and of resolution effect, 
counterpropagating spin-waves cannot be observed when scanning across
 the magnetic line (as sketched in Fig. \ref{dispAF}): a single peak 
is usually  measured  (see  Fig \ref{allqscan}a-c). A special scattering  
geometry has thus been adapted  to determine the spin-velocity with 
accuracy  in YBa$_2$Cu$_3$O$_{6.1}$\cite{shamoto} and in three different 
monolayer undoped cuprates(La$_2$CuO$_4$, Nd$_2$CuO$_4$ and  Pr$_2$CuO$_4$)
\cite{jin}. The deduced in-plane antiferromagnetic superexchange 
coupling  $J$ is typically 130 meV (Fig. \ref{figJfcucu}).  
However, $J$ does  not exhibit a monotonous behavior versus the bonding 
Cu-O-Cu length  likely on account of detailed structure of 
each system. It underlines  that the large enhancement of $J$ is caused by 
other structural units like  the Cu-O-O triangle\cite{eskes}. 

\begin{figure}
\parbox{7.5 cm} {
\epsfig{file=figja.epsi,height=8 cm,width=7 cm}
\caption[figJfcucu]{\label{figJfcucu}In-plane superexchange 
interaction determined by INS experiments versus Cu-Cu 
distance in different cuprates (from \cite{jin}).}  } 
\hspace{ .2 cm}
\parbox{7.5 cm} {
\epsfig{file=magnon_opt.epsi,height=7 cm,width=7 cm}
\caption[optmagnon]{\label{optmagnon}Energy dependence of the neutron intensity 
at $q_{AF}$ from spin-wave scattering at T=10 K (from \cite{dmitry}).
The closed (open) symbols represent the intensity of the optical 
(acoustic) excitations. The full line is a fit by a step-like function.} }
 \end{figure} 

Further, the unit cell of YBCO contains pairs of closely spaced 
CuO$_2$ layers, the bilayers. The intrabilayer coupling, referred
as $J_{\perp}$, removes the degeneracy between even- and odd-parity
electronic states. In the AF N\'eel state, these excitations
correspond to optical (dashed line in Fig. \ref{dispAF}) and acoustic 
(full line in Fig. \ref{dispAF}) spin waves, respectively.
These two modes display complementary dynamical structure factors
along the momentum transfer perpendicular to the basal plane, $q_l$:
$\sin^2(\pi z q_l)$ for the acoustic mode and $\cos^2(\pi z q_l)$
for the optical mode. (Here, $z=0.29$ is the reduced distance 
between nearest-neighbor Cu spins within one bilayer).  
This allowed to distinguish these two excitations
and to determine the optical gap\cite{dmitry,hayden2}. Fig. 
\ref{optmagnon} show the energy dependence of the neutron intensity at 
$q_{AF}$ obtained from our results\cite{dmitry} which gives
an optical magnon gap at $\omega_{opt}= 67\pm 5$ meV. 
Hayden {\it et al}\cite{hayden2} have reported a slightly larger
value from less accurate data. 
Detailed spin-wave calculations\cite{rossat2,shamoto} reveal an 
optical magnon gap at\cite{dmitry}, $\omega_{opt}=2 \sqrt{J_{\perp}J}$.
Using the value of $J=120$ meV\cite{shamoto,hayden2} for the 
in-plane superexchange, one deduces $J_{\perp}= 9.6$ meV. However, 
the above relation does not account from quantum corrections of the
AF ground state. A more accurate treatment using Schwinger bosons
representations\cite{mimo}, gives $J_{\perp}\sim 12$ meV 
which is in good agreement with band theory predictions\cite{band}, 
$J_{\perp}\sim 13$ meV. 
In classical superexchange magnetic theory, where $J$ is  
proportional to the square of the overlap of the electronic wavefunctions,
one can deduce the ratio between the intrabilayer and the inplane hopping 
matrix elements as $\frac{t_{\perp}}{t}=\sqrt{\frac{J_{\perp}}{J}}=0.34$.
This non-negligible ratio, which unlikely would vary with doping, 
shows that the electron transfer processes between direcly adjacent 
layers could play an important role in the high-$T_c$ mechanism as 
it was suggested in the interlayer tunneling model\cite{interlayer}.

Furthermore, by calibration with phonon cross section, one can determine 
 the spectral weight of the spin susceptibility in 
absolute units. Surprisingly,  the spin wave spectral weight is found 
smaller than expected from quantum corrections\cite{jin}.
This reduction of about 30\%  is presumably due to covalent effects
between copper d-orbitals and  oxygen p-orbitals\cite{covalence}. 
Reducing the absolute scale of the atomic form factor, such effects can 
also explain the reduction of the low temperature ordered magnetization 
value\cite{rossat2,covalence}.

\vskip 1 cm
\noindent
{\bf 5. WAVE VECTOR DEPENDENCES IN THE METALLIC STATE}
\vskip .5 cm

We now turn to the results in the metallic state of cuprates where 
the magnetic scattering is only found  inelastic, corresponding to dynamical 
fluctuations and peaked around the AF wave vector. The existence of 
these fluctuations is already surprising 
since in usual simple metals fluctuations arising from free electrons are 
too weak to be observed. This observation then signs the existence of 
strong electronic correlations in high-$T_c$ cuprates. 

\vskip .5 cm
\noindent
{\bf Even and Odd excitations in YBCO: $q_l$-modulation in YBCO}
\vskip .5 cm

In conventional band theory, interactions within a 
${\rm CuO_2}$-bilayer yield bonding and antibonding bands in the 
metallic state. Transitions between electronic states 
of the same type (bonding-to-bonding or antibonding-to-antibonding) 
and those of opposite types are characterized by even or odd symmetry, 
respectively, under exchange of two adjacent $\rm CuO_2$ layers. 
As a result, odd and even excitations then exhibit a structure
factor along $q_l$ similar to the acoustic and optical spin-wave 
in the N\'eel state, respectively.  As discussed in the previous section, 
this yields a $\sin^2$-type $q_l$-dependence of the lower energy 
excitations, the odd excitations. This structure factor has been effectively 
observed in any low energy (below $\sim$ 45 meV) magnetic studies. 
It is, for instance, the case in Fig. \ref{qlscan} in the SC state of 
optimally doped YBCO. Similar results have been reported in the normal 
state as well\cite{rossat2,tranquadaprb,jltp}. In weakly doped 
metallic state, $x\sim 0.5$, this modulation actually occurs because 
even excitations exhibit a gap around 53 meV\cite{bourges}, which is
reminiscent of the optical magnon gap\cite{dmitry}. At higher doping, 
the even excitations gap is lowered further by doping ($\sim$ 35 meV in 
$x\simeq 0.7$)\cite{tony4}. Surprisingly, the sin$^2$-structure factor is 
still observed at energies above this even gap (Fig. \ref{qlscan}). However, 
Fig. \ref{qlscan} suggests that this modulation is not complete as 
even excitations are sizeable at $q_l \simeq 3.5, 7$ but with a magnitude 
$\sim$ 5 times smaller. Therefore, although the even excitations occur
in the same energy range, they are surprisingly much weaker in amplitude 
than the odd excitations. Moreover,  even excitations display unexpected 
temperature dependences\cite{bourges,tony4} 
as they are strongly reduced going fom 5 K to 200 K. 

\begin{figure}
\parbox{7.5 cm} {
 \epsfig{file=ql692_40mev.epsi,height=6 cm,width=7 cm}
\caption[qlscan]{\label{qlscan} $q_l$-scan at $E= 40$ meV
in YBCO$_{6.92}$ displaying  a modulation  typical of odd
excitation (from \cite{rossat2}).
The background, obtained from q-scans across the magnetic line
(open squares), is represented by the  dashed line. The full line 
correspond to a fit by $a + b F^2(Q)\sin^2 ( \pi z q_l)$ 
above the background and where $F(Q)$
is the Cu magnetic form factor.} 
}
\hspace{ .2 cm}
\parbox{7.5 cm} {
 \epsfig{file=inpla_bz.epsi,height=6 cm,width=7 cm}
\caption[inplabz]{\label{inplabz} In-plane Brillouin zone. The shaded
area sketches the location of AF fluctuations. The closed circles 
sketch the four-peaks magnetic scattering observed at low energy in 
the LSCO system at $Q=(\pi(1\pm\delta),\pi)$ 
and $Q=(\pi,\pi(1\pm\delta))$, displaced from the 
AF momentum by an amount, $\delta= 0.245= 0.28$ \AA$^{-1}$ 
for $x=0.14$\cite{inclsco}. }
}
 \end{figure} 

\vskip .5 cm
\noindent
{\bf Commensurate fluctuations in YBCO}
\vskip .5 cm

We then discuss the in-plane wave vector dependence observed at low 
energy results, i.e. below $E\sim 40$ meV. As shown in Fig. \ref{allqscan}, 
the magnetic scattering is found in YBCO peaked at the AF momentum, 
$(\pi,\pi)$, at any doping. However, it has been recently 
reported\cite{incdai} that the magnetic fluctuations in an underdoped 
YBCO$_{6.6}$ ($T_c=63$ K) sample would become incommensurate around 
$E\sim 25$ meV, but only below $T\sim$ 70 K. This contradictory 
result needs to be confirmed as only a few measurements\cite{sternlieb} 
have previously indicated, at most, a flat-topped shape of low-energy 
q-scans profiles. In any case, typical observed peaks are far from 
simple sharp Lorentzian-shape peaks. As shown for instance at $E=$
17 meV in YBCO$_{6.83}$ in Fig. \ref{allqscan}h, two broad peaks 
could better described the observed profile. Actually, the extension in 
q-space of the magnetic scattering is quite important as its 
q-width, defined as the Full Width at Half Maximun (FWHM) $\Delta_q$, 
is typically about one fifth of the Brillouin zone size. 
On increasing doping, the peak broadens (see Fig. \ref{allqscan}), 
giving at most $\Delta_q \simeq 0.45$ \AA$^{-1}$\cite{sympo,lpr} 
(after resolution deconvolution when fitting by a Gaussian shape) 
for highly doped samples, $x\ge 0.9$ whereas $\Delta_q \sim 0.25$ 
\AA$^{-1}$ is found for weakly doped samples, $x\sim 0.5$.
This q-width may  be associated with a length in real space, 
$\xi=2/\Delta_q$, which would correspond to an AF correlation length in
a localized spins picture. Typically,
this length is found very short\cite{sympo}, as $\xi/a \sim 1-2.5$ 
depending on doping (where a=3.85 \AA). Furthermore, $\Delta_q$, still 
obtained by a Gaussian fitting of the q-lineshape, 
displays no temperature dependence within error bars (see e.g. \cite{lpr}).

\vskip .5 cm
\noindent
{\bf Incommensurate fluctuations in LSCO}
\vskip .5 cm

In LSCO, the low energy fluctuations clearly differ from what is observed 
in YBCO: a four-peaks structure is observed\cite{inclsco,sylv} instead 
of a broad peak centered at $(\pi,\pi)$ (Fig. \ref{inplabz}). 
Each spin scattering peak occurs at an incommensurate wave vector
which increases with  doping. It exhibits a q-width much smaller than 
in YBCO such as the four peaks do not overlap. 
  On increasing energy, the peaks broaden and 
concomitantly the incommensurability disappears: the spin scattering 
being maximum at $(\pi,\pi)$ with a large q-width for energies above 
$\sim$ 25 meV for $x=0.15$\cite{sylv}.


The origin of the incommensurability in LSCO has been widely discussed 
in the framework of itinerant magnetism\cite{levin,theoinc,little,nest}. 
Magnetic correlations are enhanced at the incommensurate wave vectors 
due to a near nesting property of the simple bidimensional Fermi 
surface deduced from simple tight-binding calculations. At the 
AF wave vector, low energy spin fluctuations are removed up to 
an energy which is twice the chemical potential\cite{nest}. This
picture seems to nicely account for the observed features. However, the 
calculation of the chemical potential from the doping level cannot be 
deduced from conventional band structure calculations\cite{little} and 
would be associated with new quasi-particles charge carriers\cite{ops}. 
This model can also explain why the LSCO and YBCO systems display 
different momentum dependences according to a different Fermi 
surface topology in the two systems\cite{levin,theoinc}. However, 
quite recently, 
another interpretation, based on real-space domains observed in insulating 
phases caused by charge ordering forming stripes\cite{stripes}, 
has been proposed to explain the  discommensuration\cite{emery}.

\vskip 1 cm
\noindent
{\bf  6. ENERGY DEPENDENCE OF THE SPIN SUSCEPTIBILITY}
\vskip .5 cm

\begin{figure}
\parbox{7.5 cm} {
\epsfig{file=chisc.epsi,width=8 cm, height= 20 cm}
\caption[suscep_sc]{Imaginary part of the odd spin susceptibility at T= 5 K 
in the superconducting state for six oxygen contents in YBCO. 
At the top, the spin waves scattering of the undoped parent compound at 
T=8 K is displayed for comparison. 
 (See Fig. \ref{suscep_ns} for other details). }
\label{suscep_sc} 
}
\hspace{ .2 cm}
\parbox{7.5 cm} {
 \epsfig{file=chi_100k.epsi, width=8 cm, height= 16.5 cm}
\caption[suscep_ns]{Imaginary part of the odd spin susceptibility at 
T= 100 K in the normal state for four oxygen contents in YBCO. 
 These curves are 
directly comparable each others from measurements  performed on the same
triple axis spectrometer (2T-Saclay). They have been normalized to the 
same units using standard phonon calibration\cite{tony2}.
Using further measurements\cite{bourges}, 100 counts in the vertical scale 
would roughly correspond to $\chi_{max} \sim$ 350 $\mu_B^2$/eV 
in absolute units. These curves are in the same units as those of Fig. 
\ref{suscep_sc}. Lines are fit by Eq. \ref{lor}.
\label{suscep_ns} 
}  }
 \end{figure} 

The most remarkable INS result in high-$T_c$ cuprates is certainly 
the drastic change of the spin dynamics when entering  the superconducting 
state. Basically, a gap opens at low energy, $E_G$, and a strong resonance 
peak appears in the odd spin susceptibility at a characteristic energy, $E_r$.
These features are most likely related to coherence effects 
because of spin pairing of the superconducting electron  pairs as it has been 
inferred in conventional BCS formalism\cite{schrieffer}.
However, the observed characteristics imply unconventional 
superconducting gap symmetry, i.e. anisotropic gap in {\bf k}-space, 
most probably of d-wave symmetry. These features as well as their 
implications are described in the two next sections.
The normal state properties will be discussed in a third section.

\vskip .5 cm
\noindent
{\bf doping dependence of the odd spin susceptibility and resonance peak}
\vskip .5 cm

In order to emphasize these features, it is convenient to discuss the 
energy shape of the odd spin susceptibility. Fig. \ref{suscep_sc} depicts
$Im \chi$ in the superconducting state for energies below 50 meV and 
at the AF wave vector: $Im \chi$ is displayed for 6 different 
oxygen contents, 4 in the underdoped regime and 2
in the overdoped regime. The normal state (T= 100 K) spin susceptibility is 
reported on Fig. \ref{suscep_ns} for four different samples equivalent 
to those reported in the SC state: 3 in the underdoped regime, 1 in the 
overdoped regime. All these measurements have been scaled to the same 
units from an analysis as discussed in section 3. The results have been 
obtained on the same sample, except for the 
$x=0.97$ sample\cite{hoechst} which have been scaled using phonon 
calibration.

One first clearly notices that $Im \chi$ strongly evolves with doping. 
$Im \chi$ is characterized by a maximum which becomes sharper in energy 
for higher doping. In addition to that peak, magnetic correlations occur 
at other energies. For instance,  the spin response in the nearly optimally 
case YBCO$_{6.92}$ displays a strong  resolution-limited peak located
 at 41 meV and a plateau 
 in the range 30-37 meV. Normal state AF fluctuations exist in a
wider energy range with a broad maximum around 30 meV.  The  41 meV 
enhancement has totally vanished at T= 100 K and has been, therefore, 
assigned to a ``magnetic resonance peak'', $E_r$\cite{rossat1}.

Upon increasing doping, the resonance peak is slighly 
renormalized to lower energies whereas the plateau is substantially 
removed, so that, the SC spin susceptibility is basically well accounted for by 
a single sharp peak around $E_r$= 39-40 meV for fully oxidized YBCO$_7$ 
samples. A clear agreement is now established among the neutron community
about this slightly overdoped regime \cite{hoechst,tony1,mook,lpr}.
The most important result is that this resonance peak intensity disappears 
at $T_c$  as shown in Fig. \ref{resotemp}. Performing a q-scan at the 
resonance energy in both the SC state and in the normal state emphasizes 
the vanishing of the resonance peak above $T_c$: Fig. \ref{qscan39} displays 
a q-scan at $\hbar\omega=$ 39 meV with very weak magnetic intensity 
at $(\pi,\pi)$ and T= 125 K\cite{jltp}.
This striking temperature dependence demonstrates that this 
peak is intimately related to the establishement of superconductivity.  
Fig. \ref{resotemp} shows that the resonance intensity actually follows 
an order parameter-like behavior\cite{rossat1,hoechst,mook,tony2}
whereas the resonance energy is itself very weakly temperature 
dependent, if any\cite{hoechst,dai,tony2}. The q-scan at 39 meV also underlines 
the strong reduction of the normal state  AF correlations in the  overdoped regime. This is in sharp contrast with the nearly optimally doped 
sample, YBCO$_{6.92}$\cite{lpr}: on a more quantitative ground, 
Fig. \ref{suscep_ns} reveals that the AF correlations amplitude is reduced 
at T=100 K by a factor $\sim$ 2-4 (depending of the energy) going 
from $x=0.92$ to $x=1$.    

On lowering doping, the spin susceptibility spreads out over a 
wide energy range. However, a close inspection of {\bf q}- and temperature 
dependences of $Im\chi$ reveals that the resonance peak feature is still 
present in underdoped samples\cite{sympo,epl,lpr}. This can be 
demonstrated by making the difference between the neutron intensity 
measured at T= 5 K at Q=$(\pi,\pi)$ and the same scan measured just 
above $T_c$\cite{dai,tony3}: this difference exhibits a sharp peak 
at energy where $Im\chi$ at T=5 K is maximum. Here, the whole lineshape of
$Im \chi$ is preferably reported as it can be directly compared 
with theoritical models. The resonance peak energy 
is shifted to lower energy in underdoped regime,  actually following 
$T_c$\cite{epl,tony3} (Fig. \ref{ertc}). The relative amplitude
of the resonance peak as compared to the normal state intensity
is decreasing with lowering oxygen content, so that, the resonance
amplitude is about 50\% of the maximum intensity for heavily doped 
samples ($x\ge$ 0.6, $T_c > 60$ K , when the resonance energy is located 
around $\sim$ 34 meV, )\cite{dai,epl,tony4}, and only about 15\% for 
weakly-doped samples ($x\simeq$ 0.5, $T_c \sim 50$ K and $E_r\simeq$ 25 
meV)\cite{bourges}. In contrast, the intensity at the maximum at 
T= 5 K actually increases
with decreasing doping. Conversely, the normal state excitations 
is larger in the weakly-doped range (see Fig. \ref{suscep_ns}).
Therefore, although the spectral weight of $Im\chi$ becomes larger, 
the resonance feature vanishes for smaller doping and $T_c$.

\vskip .5 cm
\noindent
{\bf Spin gap in the SC state }
\vskip .5 cm

In the superconducting state, the spin susceptibility is limited 
at low energy by a gap below which no AF scattering is visible.
This energy gap is then defined by the first inflexion point of 
$Im\chi$ curve and is referred to as ''spin-gap``, $E_G$. Below $E_G$, 
q-scans display no or very weak peak at $(\pi,\pi)$, as displayed, 
at 10 meV  in Fig. \ref{allqscan}g and Fig. \ref{allqscan}j. 
Going into the normal state (T=100 K), the spin-gap is suppressed because low 
energy excitations are sizeable at $(\pi,\pi)$. Moreover,  
the peak intensity appears on heating at $T_c$\cite{lpr} demonstrating 
that this spin gap is directly related to the superconducting gap.
Further, $E_G$ increases upon  doping\cite{sympo} but with a 
different dependence than the resonance energy. 

In most  cases, $E_G$ is defined by a sharp resolution-limited 
step (Fig. \ref{suscep_sc}). However, this low-energy step is found 
much broader for $x=$ 0.83. Concomitantly, 
the superconducting transition of this sample was also broader than usual.
Therefore, it is reasonable to attribute this behavior to the lack of 
 sample homogeneity which could smear the sharp features of the spin response. 
Nevertheless, a resonance peak has been clearly observed in that 
sample\cite{sympo} whose relative amplitude might be only reduced 
as compared with subsequent reports\cite{dai,epl,tony3}.
 
This comparison between spin-gap and $T_C$-width also gives some insight
about two related issues. First, in the weakly-doped YBCO$_{6.5}$ regimes,
this spin-gap has been widely reported by the french 
group\cite{rossat2,rossat1,sympo,lpr} and not by others\cite{tranquada} for
sample having similar oxygen contents and $T_c$. As a matter of fact, a close 
inspection of the $T_c$-curves reveals that the samples exhibiting a 
spin-gap had clearly a sharper SC transition. Second, it may solve a 
controversy existing in LSCO system where a spin-gap has been reported 
only recently\cite{sylv,yamada}. Previous studies have not succeeded to 
detect it\cite{inclsco} likely because of sample inhomogeneities and/or 
impurity effects. More generally, the observation
of the spin-gap appears to be very sensitive to sample defects and 
impurities. As an example, the controlled substitution of 2\% of zinc in 
overdoped YBCO (where $T_c$ is reduced by $\sim$ 20 K) induces low energy 
AF fluctuations below the spin-gap\cite{yvan,sidis}. 

Finally, it is very interesting to draw a parallel between the 
spin response in superconducting state of LSCO near maximum $T_C$ 
($x=$ 0.14) and in weakly-doped underdoped YBCO$_{6.5}$.
Both systems display a small spin gap in the SC state
of about 2-5 meV. Both systems also exhibit a subtle enhancement of the 
spin susceptibility just above the spin gap: in LSCO, it has been 
underlined\cite{mason} that around 9 meV an additional enhancement occurs 
only in the SC state which is accompanied to a sharpening in momentum space.
Similar observations have been previously reported around 7 meV in 
YBCO$_{6.52}$\cite{rossat3}. In light of the recent experiments in 
YBCO$_{6.5}$\cite{tony3}, it seems that this effect is different 
from the resonance peak feature occuring at higher energy. However, this 
issue is still under debate as no resonance peak has been detected so far 
in LSCO. Besides, this could be simply because the resonance intensity is 
expected to be only $\sim$ 10 \% of the total magnetic scattering for such low 
$T_c \le 40$ K, still by comparison with YBCO$_{6.5}$\cite{bourges}.

\begin{figure}
\parbox{7.5 cm} {
\epsfig{file=ertc.epsi,height=7 cm,width=7 cm}
\caption[ertc]{\label{ertc} Resonance energy in YBCO versus $T_c$.
French results are from \cite{hoechst,sympo,epl,lpr}, Princeton results 
are from \cite{tony1,tony3}, Oak Ridge results are from \cite{dai,mook}.} 
 }
\hspace{ .2 cm}
\parbox{7.5 cm} {
\epsfig{file=sqw_10mev_ft.epsi,height=7 cm,width=7 cm}
\caption[sqw683]{\label{sqw683} Temperature dependence of the peak 
intensity at low energy in YBCO$_{6.83}$. Background has been subtracted 
using q-scans (see Fig. \ref{allqscan}). }
 } 
 \end{figure} 

\vskip 1 cm
\noindent
{\bf  7. COHERENCE EFFECTS IN THE SUPERCONDUCTING  STATE}
\vskip .5 cm

As the resonance peak and the spin-gap are both intimitaly related to 
the superconductivity, a simple 
interpretation has then been proposed in itinerant magnetism 
approaches\cite{mimo,shiba,reso_wces,scgap+vhs,flora,reso_flora,levinreso,yakovenko,pwa}.
Beyond different hypotheses, these models basically show that the 
resonance peak occurs simply because of the BCS pairing. 
Starting from a Fermi liquid picture, the non-interacting electronic 
spin susceptibility (Lindhard Function) is written as,

\begin{equation}
\chi^{\circ}(q,\omega)= (g \mu_B)^2 \lim_{\epsilon \rightarrow 0} 
\sum_k {{f_{q+k}-f_{k}}\over
 {\epsilon_{q+k}-\epsilon_{k} -\hbar\omega -i\epsilon}}
\label{lindhard}
\end{equation}

where $\epsilon_{k}$ is the electronic band dispersion 
and $f_{k}$ is the associated Fermi function. This spin susceptibility  
is described by the two-particle response function which usually 
gives featureless broad response due to sum over the reciprocal space.
However, band structure singularities and nesting effects can induce 
rather complex lineshapes and q-dependences.
Going into the SC state, one should account for the spin pairing 
of Cooper pairs\cite{schrieffer}, $\chi^{\circ}$ then becomes for T=0,

\begin{equation}
\chi^{\circ}(q,\omega)= (g \mu_B)^2  \lim_{\epsilon \rightarrow 0} 
\sum_k { \Big[ {1-{{\Delta_k \Delta_{q+k} + 
  \epsilon_{q+k} \epsilon_{k}  }
\over{ E_{q+k} E_{k} } } }  } \Big]
{{1-f_{q+k}-f_{k}}\over
 {E_{q+k}+E_{k} -\hbar\omega -i\epsilon}}
\label{bcs}
\end{equation}

with $E_{k}=\sqrt{\Delta_k^2+\epsilon_k^2}$ and $\Delta_k$ is the 
{\bf k}-dependent
SC energy gap. The term in brackets is a coherence factor related to the BCS 
pairing\cite{schrieffer}. This function displays a soft edge behavior 
when the sum $(E_{q+k}+E_{k})$ is mimimum\cite{flora,pwa} but not a sharp 
peak. Further, the q-dependence of this function is not necessarily 
peaked around $(\pi,\pi)$\cite{yakovenko}.
Therefore, neutron data cannot be reproduced in a generic noninteracting 
electron model. To overcome these difficulties, 
a Stoner-like factor should be included which is related either to 
band structure singularities, to spin fluctuations, or even
to interlayer pair tunneling effects\cite{pwa}. In the case 
of a  magnetic exchange, $J(q)$,  the interacting spin susceptibility 
using  Random Phase Approximation (RPA), is
\begin{equation}
\chi(q,\omega)=\frac{\chi^{\circ}(q,\omega)}{1-J(q) \chi^{\circ}(q,\omega)}
\label{rpa}
\end{equation}

The RPA treatment have been held responsible for the 
sharpness of the peak in both energy and momentum. 
Using this expression, a sharp peak is obtained above a gap and both
are  proportional to the SC gap. In the normal 
state, featureless non interacting spin susceptibility 
for itinerant carriers is restored. Hence, this result in both the normal 
state and the SC state agrees with what is measured in the overdoped 
regime.  

In this view, the resonance energy, $E_r$, closely reflects 
the amplitude of the SC gap as well as its doping dependence.
 Experimentally,  this prediction is supported as 
the resonance energy is found to scale with $T_c$ 
with $E_r/k_B T_c \simeq 5.1$ (Fig. \ref{ertc}). However, the true 
proportionality of the resonance energy with the maximum SC energy gap,
$\Delta_{SC}^{max}$, is not simply 2 as expected in simple BCS theory with 
an isotropic SC gap. 
Here, it crucially depends on band structure effect\cite{ops}.
For the same reason, this relation is even more complex for the spin gap
doping dependence\cite{flora}. Interestingly, this model requires 
d$_{x^2-y^2}$-wave symmetry of the SC gap function,
$\Delta_k$\cite{shiba,reso_wces,scgap+vhs,flora,reso_flora,levinreso},
as a change of sign of $\Delta_k \Delta_{Q_{AF}+k}$ should occur  in the coherence factor.
However, other subtle scenario can occur\cite{yakovenko}.
Van-Hove singularities in band structure\cite{scgap+vhs,reso_flora} 
have been also invoked to play a role in the peak enhancement. 
As the peak position of the resonance exhibits very little temperature 
dependence\cite{hoechst,dai,tony2}, the SC gap should not change with temperature in this framework. 

Finally, one can conclude that coherence effects related to the spin 
pairing seem to account for the marked modifications 
of the spin susceptibility in the SC state. Moreover, the additional 
enhancement in the SC state observed $\simeq 7-9$ meV\cite{mason,rossat3} 
can be also explained in this framework: a transfer of the spectral 
weight from the low energy (below the spin-gap,  $E_G$) to energies just 
above $E_G$ is indeed expected. In any case, the major limitation 
of this model is that in the underdoped regime (as well as for optimal 
doping) one observes another magnetic contribution existing in the SC 
state and remaining in the normal state.  This more complex 
behavior of the spin susceptibility requires more sophisticated models
than simple itinerant hole picture.

A correlated-electron model which is aiming to unify superconductivity  
and antifer\- romagnetism\cite{so5}, has been recently proposed to account 
for the resonance peak. A particle-particle collective excitation occurs 
in this model\cite{demler}, whose matrix element coupling this excitation
to the magnetic neutron cross section, vanishes in the normal state, but is nonzero in the superconducting state.  In this model, because
excitation energy is proportional to the hole doping\cite{demler}, 
the resonance energy should increase with the oxygen content. 
Experimentally, this trend is found up to optimal doping.
However, going from YBCO$_{6.9}$ to YBCO$_{7}$, $E_r$ is rather reduced 
(Fig. \ref{suscep_sc}) following actually $T_c$  in contrast to the 
expectation.

\vskip 1 cm
\noindent
{\bf 8. SPIN PSEUDO-GAP AND QUASI-DISPERSION}
\vskip .5 cm

As stressed in previous sections, the resonance feature is accompagnied 
in underdoped and optimally doped regimes by another broad contribution
which is rapidly stronger in amplitude at lower doping.
The odd spin susceptibility is actually  maximum at a characteristic 
energy, $\sim$ 30 meV (see Fig. \ref{suscep_ns}). This maximum 
of $Im\chi$, which directly corresponds to the pole of the spin 
susceptibility, then naturally defines a gap in the spin excitation spectrum. 
Obviously, this gap can  be identified to the well-known  ``spin 
pseudo-gap''\cite{rmn,rossat1,lpr} although its definition is different 
from previous convention\cite{rossat1}. 

For the sake of clarity, it is useful to describe the normal state 
spin susceptibility   by a damped Lorenztian function.  This expression 
is usually applied to disordered short range localized spins. It is, 
for instance, the case of one dimensional magnetic system  such as 
the so-called Haldane gap systems\cite{NENP}. $Im\chi$ is then written,

\begin{equation}
Im \chi (q,\omega) = 
     {{\chi_q \ \ \omega \ \ \Gamma}\over{ (\omega-\omega_{pg})^2+\Gamma^2}}
\label{lor}
\end{equation}

\noindent 
where $\omega_{pg}$ is the characteristic energy corresponding to the pole 
of the spin susceptibility. Eq. \ref{lor} is giving a good description of 
the normal state spin susceptibility (Fig \ref{suscep_ns}) with
$\omega_{pg}\simeq 28$  meV  in heavily-doped samples ($x\ge 0.6$).
 In that doping range, the spin pseudo-gap, $\omega_{pg}$, has no marked 
doping as well as temperature dependences.  In contrast 
the spin gap, defined in Sec. 6, increases with doping. Therefore,  
these two gaps clearly differ. However, these two gaps ``accidentally'' 
occur in the same energy range yielding complex lineshapes in the SC state. 
Quite independently of the doping, the excitations display a strong damping 
with $\Gamma \simeq$ 12 meV.  As a result, sizeable excitations exist in the 
normal state down to low energy. This differs from the SC state where
the low energy excitations are much more reduced by the spin-gap. 
Fig. \ref{sqw683} displays the temperature dependence of the magnetic 
scattering at $(\pi,\pi)$ and at $\hbar\omega$= 10 meV for $x=0.83$. In the 
SC state, small residual magnetic scattering occurs at this energy for that 
sample (Fig. \ref{allqscan}). Upon heating, the AF fluctuations increase 
in the normal state passing through a maximum around a temperature 
$T^*\sim$ 120 K larger than $T_c$. This temperature behavior reminds of
that measured by copper NMR experiments in the underdoped 
regime\cite{rmn,rmnhere} where the spin-lattice relaxation rate $^{63}T_1$
is related to the spin susceptibility by $1/^{63}T_1T\propto \sum_q 
Im\chi(q,\omega)/\omega$. Similarly to Fig. \ref{sqw683}, 
$1/^{63}T_1T$ displays a maximum at a temperature $T^*$ in underdoped 
cuprates. This result has been widely interpreted by the opening of a 
spin pseudo-gap at the AF wave vector below $T^*$\cite{rmn,rmnhere}. 
More probably, this unusual temperature behavior as well as the value of 
$T^*$ rather results from the interplay of the magnetic parameters, 
namely $\omega_{pg}$, $\Gamma$ and $\chi_q$, which have different 
temperature dependences.

In weakly-doped samples, YBCO$_{6.5}$, $\omega_{pg}$ is located $\sim 20$ meV 
at T= 100 K (Fig. \ref{suscep_ns}); $\omega_{pg}$ is also temperature 
dependent\cite{lpr,bourges} reaching 30 meV for T$\ge$ 200 K: this trend 
is caused by the fact that the low 
energy excitations (below $\simeq 25$ meV) increase on decreasing 
temperature although the high energy part does not change\cite{lpr}.
This yields a  critical-like behavior which can be associated with  
the proximity of the AF quantum critical point. Furthermore, similar energy 
and temperature dependences are found for samples
in the insulating phase nearby $x\simeq 0.4$\cite{yvan,sendai}.

\vskip .5 cm
\noindent
{\bf Quasi-dispersion}
\vskip .5 cm

This normal contribution actually extends to higher energies and a significant 
spectral weight at energies comparable to $J$ is observed\cite{bourges,hayden}.
Furthermore, it has been recently reported in YBCO$_{6.5}$ that 
the high energy spin excitations exhibit a quasi-dispersion
behavior. Above $\hbar\omega \sim 50$ meV,  a double peak structure
emerges from  constant energy q-scans\cite{bourges}. This evidences 
a noticeable inplane propagating character of the spin excitations as,
at each energy, one can define 
a characteristic wave vector which varies and matches a dispersion curve
reproduced in Fig. \ref{quasidisp}. Similarly to magnons in an ordered 
magnetic system, one then observes propagating excitations in the 
metallic phase of cuprates. 

\begin{figure}
\parbox{7.5 cm} {
\epsfig{file=quasi_65.epsi,height=8 cm,width=7 cm}
\caption[quasidisp]{\label{quasidisp} Spin excitation spectrum 
for odd - ``acoustic'' -
(open circles) and even - ``optical'' - (closed circles) excitations 
at 5 K in YBCO$_{6.5}$ (from \cite{bourges}). The lowest open circle 
represents the energy of the maximum of the odd susceptibility.
The horizontal bars represent the intrinsic q-width (FWHM) after a 
Gaussian deconvolution from the spectrometer resolution. Full
lines correspond to an heuristic quadratic fit like
$\omega^2=\omega_0^2+c^2 q^2$. Dashed lines represent the magnon
dispersion curves of the undoped materials.  } }
\hspace{ .2 cm}
\parbox{7.5 cm} {
\epsfig{file=suscep_2d.epsi,height=11 cm,width=7.5 cm}
\caption[suscep2d]{\label{suscep2d} Comparison of the q-integrated spin susceptibility per ${\rm CuO_2}$ plane in YBCO$_{6.5}$\cite{bourges} 
and in ${\rm La_{1.86}Sr_{0.14}CuO_4}$\cite{hayden}. The dashed 
lines represent the 2D-integrated AF spin-wave contribution which is 
nearly constant in such an energy range (see\cite{bourges,jin} for details). }  
}
 \end{figure} 

Using classical spin-wave formalism for 
an Heisenberg model  (which in principle cannot apply to such 
non-magnetic ground-state), one can describe the observed 
dispersion as a softening of the magnon dispersion  of the undoped
materials (Fig. \ref{quasidisp}).
From a simple parabolic fit, one deduces a spin velocity,
 c $\simeq$ 420 meV \AA, 65\% of the AF spin wave 
velocity of 650 meV\AA\cite{shamoto,hayden2}. This softening can be 
itself expressed in terms of an effective AF exchange as 
$J_{eff} \approx 0.65 J^{AF} \simeq $ 80 meV at low temperature. 
A gap is found $\sim$ 55 meV for the even excitations which  
slightly increases with temperature\cite{bourges}.
This even gap in the metallic state is found to be reduced from 
the magnon optical gap, $\omega_{opt}^{AF} = 67$ meV\cite{dmitry}. 
This effect can be readily accounted for by the same effective 
AF exchange using  the classical spin-wave theory as,
 $\omega_{even}=2\sqrt{J_{eff} J_{\perp}}$. 
However, as a limitation of this approach, counter-propagating excitations 
are not better resolved in the higher energy range\cite{bourges}. 
Further, these observed peaks exhibit an intrinsic q-width which can 
be related to a real space correlation length of only about 9 \AA.
Therefore, instead of being 
well-defined propagating excitations, like magnons in the pure N\'eel 
state, one rather observes magnetic fluctuations which propagate only 
over small in-plane regions.

\vskip .5 cm
\noindent
{\bf  LSCO versus YBCO}
\vskip .5 cm

Finally, it is instructive to compare the energy dependences of the 
spin response in LSCO near optimal doping ($x \sim$ 0.15) and in 
weakly-doped underdoped YBCO$_{6.5}$. Above $\sim$ 25 meV when the 
magnetic scattering becomes commensurate in LSCO\cite{sylv}, the 
neutron  results show similar trends in the two systems. For instance, 
at about E $\sim$ 50-60 meV, the magnetic spectrum is strongly reduced 
at $(\pi,\pi)$\cite{rossat1,sympo,lpr,sylv}. This effect was previously 
attributed  to an ``high energy cut-off'' but, more likely, this
actually originates, at least in YBCO, due to the interplay of a few 
effects: first, above 
$\sim$ 40 meV the spin susceptibility spectral weight at $(\pi,\pi)$ 
effectively decreases with the energy. Moreover, $Im\chi$ displays 
a dip feature around 50 meV. Finally, the magnetic scattering strongly 
spreads out over the in-plane {\bf q}-space for energies above 50 meV 
due to the quasi-dispersion behavior (Fig. \ref{quasidisp}).  In LSCO, the 
quasi-dispersion has not been reported but, as a matter of fact, it can be 
inferred from high energy measurements for $x=$ 0.14\cite{hayden} as
i) q-scan around E $\simeq$ 100 meV exbibits significant broadening 
compatible with propagating excitations ii) a zone boundary peak 
is found as one would expect from a dispersion-like behavior. 
In any case, one clearly observes in the two systems a similar
 momentum broadening at high energy.

Actually, to emphasize the comparison of the two systems, it is more 
convenient to perform  the 
q-integration  of the spin susceptibility over the 2D in-plane wave 
vector $q_{2D}$, as $Im \chi_{2D} (\omega) = \int d {\bf q}_{\rm 2D} 
Im \chi({\bf q}_{\rm 2D},\omega) /  \int d {\bf q}_{\rm 2D}$. 
This also allows to overcome the difficulty that LSCO displays non 
overlapping incommensurate peaks below 25 meV. Further, the determination 
of $Im\chi$ in absolute units makes possible the direct comparison of LSCO 
($x=$ 0.14)\cite{hayden} and YBCO$_{6.5}$\cite{bourges}. The susceptibilities, 
calculated  per single ${\rm CuO_2}$-layer are reported in the same 
absolute units in Fig. \ref{suscep2d}.  This q-integration has 
the effect to change the shape of the spin susceptibility as it enhances the 
high energy part due to the broadening in q-space of the AF fluctuations.
 For both systems, the 
spectral weight in the metallic state is very comparable to the spin 
wave spectral weight of the undoped materials  (dashed lines in 
Fig. \ref{suscep2d}).  The spin susceptibility  
exhibit a linear behavior at low energy, and then passes through a 
maximun around 25 meV, and extends to high energy. In YBCO, one finds 
in addition a dip feature around E$\simeq$ 50 meV\cite{bourges} (whose origin 
is not  clear at present) which actually might also exist in LSCO.
Finally, apart from incommensurate peaks at low energy, both systems 
exbibit very comparable momentum and energy dependences. As discussed 
above, modifications of $Im\chi$ in the SC state also show remarkable 
similarities. That strongly suggests that these two systems have 
same kind of hole doping per $CuO_2$ plane.  In any case,
this behavior is very different from the one observed in optimally doped 
YBCO. This is strongly suggesting that LSCO($x$=0.15) does not correspond to
an optimally doped cuprate in the generic phase diagram (Fig. \ref{diag}).
 
\vskip .5 cm
\noindent
{\bf which models ?}
\vskip .5 cm

What kind of models can explain these results in the normal state ?
Some elements can be pointed out to describe at least part
of the situation. First, these excitations can be associated with short range
AF ordering of the localized copper spins. Indeed, the role played by 
localized copper spins in the metallic underdoped state is an 
important issue either for the nearly AF liquid model\cite{pines,barzykin}
of in two dimensional quantum disordered
approaches\cite{flora,millis2,tjmodel,rice}.  
Using RPA approximation, the magnetic exchange 
between copper spins $J(q)$ explains why the magnetic scattering is maximum
at $(\pi,\pi)$. Incidentally, the quasi-dispersion arises from 
$J(q)$ in this framework as heuristically discussed above. 
Further,  an origin for the spin pseudogap is given in the
t-J model\cite{flora,tjmodel}, which, similarly to what is found for 
the spin ladder compounds\cite{rice}, occurs because of the formation of 
singlet RVB-states (spinon pairing).  
Second, these observations can also support itinerant magnetic models.
As an example, it could remind the case of metallic Palladium 
where non-interacting electronic spin susceptibility is enhanced due 
to ferromagnetic interactions\cite{paramag}, leading to paramagnons 
behavior. In this approach, a dispersion-like behavior can simply 
occur due to the interplay of both band 
structure singularities and interaction renormalization.
The observation of the even gap suggests the splitting of the Fermi 
surface caused by interlayer interaction. Interestingly, the even gap 
is shifted to low energies at higher doping\cite{tony4}. Therefore, 
both dispersion curves in Fig. \ref{quasidisp} tend to collapse at 
sufficiently high doping removing bands splitting. Such a doping 
dependence leads to an interesting issue regarding the photoemission 
experiments which have reported  no ''bilayer splitting`` of the 
Fermi surface in ${\rm Bi_2Sr_2CaCu_2O_8}$\cite{campu,bilayer}

\vskip 1 cm
\noindent
{\bf  10. CONCLUSION}
\vskip .5 cm

The spin dynamics in metallic high-$T_c$ cuprates as seen by Inelastic 
Neutron Scattering has been reviewed. Due the electronic interactions 
within a ${\rm CuO_2}$-bilayer, one observes in the metallic state 
two different excitations:  odd mode at lower energy (analogous 
of acoustic magnon) and even mode always weaker in amplitude 
(analogous of optical magnon). The odd susceptibility exhibits strong 
doping dependence  which is characterized by 
two distinct contributions: a ``magnetic resonance peak'' which occurs 
only in the SC state, a normal contribution characterized by a spin 
pseudo-gap.  Going into the overdoped regime in YBCO, the magnetic 
fluctuations in the normal state are strongly reduced\cite{hoechst}. 
Incidentally, it is very interesting to notice that the vanishing of magnetic 
correlations can be identified to the disappearance of the 
anomalous properties in overdoped cuprates (Fig. \ref{diag}). 

Using simple itinerant magnetism, it has been widely proposed that the
resonance results from spin-flip charge carrier excitations across the SC 
energy gap. Such Fermi liquid approaches can also positively describe 
the normal state in the overdoped regime. It can even  explain the occurence 
of incommensurate fluctuations in LSCO due to nesting effect of the Fermi 
surface. However, other aspects of the spin dynamics in underdoped and 
optimally doped samples, and in particular, the existence of the 
spin pseudo-gap cannot be accounted for. More generally, the observation 
of peaks around  $(\pi,\pi)$ implies the existence of 
 strong electronic correlations in the metallic regime. 
 
 Other interesting properties, which have not  been addressed here, are the 
effect of impurities. Zinc substitution is known to strongly reduce 
the SC temperature without changing the hole amount in the ${\rm CuO_2}$ 
planes. Interestingly, zinc also strongly modifies the spin excitation 
spectrum\cite{yvan,sidis,kakurai,lt21}: with only 2 \% of zinc, the 
resonance peak intensity is strongly reduced  and  low energy 
AF excitations appear where nothing was measurable in zinc-free 
samples\cite{yvan,sidis}. Most likely, these results suggest that zinc 
enhances the AF correlations and induces a localization of the charge 
carriers.

Finally, Inelastic Neutron Scattering experiments have 
evidenced  unusual spin dynamics in metallic cuprates which shed new light
on the strong electronic correlations in these materials. 
Nevertheless, the way the copper spins are intrinsically coupled to the 
holes remains a question under discussion and yields a very interesting 
problem of localized-itinerant duality magnetism. For sure, quantitative 
comparison of the spin susceptibility measured by NMR and INS is needed 
to understand that issue. 

\vskip .4 cm

\noindent {\bf Acknowledgments}\\

The work reviewed here is the fruit of the collaboration of many  people 
whose names appear all along the references.
Here, I  want particularly to acknowledge my close collaborators: L.P. 
Regnault (CENG-Grenoble), Y. Sidis (LLB-Saclay), B. Hennion (LLB-Saclay),
H.F. Fong (Princeton University), B. Keimer (Princeton University),
H. Casalta (ILL-Grenoble) and A.S. Ivanov (ILL-Grenoble).

 \end{document}